 \documentclass[12pt,preprint]{aastex} 

\usepackage{graphicx}

\slugcomment{Draft 2012  March 16, revised }

\shorttitle{A New Fast SiPM Photometer\altaffilmark{1} }
\shortauthors{F.Meddi et al et al.}

\begin{document}


\title{ A New Fast Silicon Photomultiplier Photometer  \altaffilmark{*}} 


\author{F. Meddi\altaffilmark{1}, F. Ambrosino, R. Nesci,  C. Rossi, S. Sclavi}
 \affil{Dipartimento di Fisica, Universit\`a La Sapienza, Roma, Italy}
\author {I. Bruni }
  \affil{INAF, Osservatorio Astronomico di Bologna, Italy}
  \author{ A. Ruggieri,  S. Sestito}
 \affil{ INFN- Sezione Roma1, Italy }
 
 \altaffiltext{*} { Based on observations made with the 152 cm  Cassini telescope at the 
  Loiano station of the Bologna Observatory and with the 50 cm telescope of 
  the Roma University La Sapienza at Vallinfreda (Roma).}  
 \altaffiltext{1}{email: franco.meddi@roma1.infn.it}

\begin{abstract}
The realization of low-cost instruments with high technical performance  is a goal which deserves some efforts in an epoch of fast technological 
   developments: indeed such instruments can be easily reproduced and   therefore allow to open new research programs in several Observatories.  
  We  realized a fast optical photometer based on the SiPM technology,   using commercially available modules. 
  Using low-cost components we have developed a custom electronic chain  to extract the signal produced by a commercial MPPC 
  module produced by  Hamamatsu, in order to obtain sub millisecond sampling of the light curve of astronomical sources, typically pulsars.
	   In the early February 2011 we observed the Crab Pulsar  at the Cassini telescope  with our prototype photometer, deriving its period, power spectrum and  shape of its light curve in very good agreement with the results obtained in the past with other instruments. 
\end{abstract}





\section{Introduction}

Astronomical sources with fast variability are basically of three kinds: pulsars, interactive binaries and pulsating stars. Many of these objects are also X-ray and Gamma-ray sources and their study is of great interest because several orbiting X-ray and Gamma-ray observatories are presently operative.  Time scales variabilities range from hours to thousandths  of seconds: amplitude variations in the optical band range from 100\% (Pulsars) down to less than 0.1\% (O Subdwarfs).  For fast time scales the only detectors available in the optical band were the classical photomultipliers. In recent times a new class of detectors has been developed,  the Silicon Photo Multipliers (SiPM), whose astronomical use is still to be explored in details. To this purpose we have built a prototype of fast astronomical photometer, based on SiPM detectors commercially available from the  Hamamtsu firm \footnote{http://www.hamamatsu.com/}.

In this work  we report the technical details of our instrument and  the first astronomical results.

\section{Technical description} \label {techdes}

Fast astronomical photometers based on new technology detectors are presently used by a  limited number of research groups:  for instance the OPTIMA  team of the Max Planck Institute (Kanbak 2003), the AQUEYE/IQUEYE team of the Padova University (Barbieri et al. 2009, Naletto et al. 2009), the S-Cam device (Oosterbroek et al. 2008), the Ultracam device (Dhillon 2008), the GASP device  (Collins et al. 2008) .

The typical characteristics  of the SiPM detectors used for our photometer  are the short response time, the segmentation in cells of linear size from 25 $\mu$m  to 100 $\mu$m, a Photon Detection Efficiency (PDE) up to 75\% at 450 nm. 
 The sensor  present inside the MPPC (Multi Pixel Photon Counter) modules
is the S10362-11-050U, whose main characteristics are reported in the Hamamatsu web page. 
  \footnote{http://sales.hamamatsu.com/assets/pdf/parts\_S/mppc\_kapd0002e08.pdf} 
 
\begin{figure*}
\centering
 \includegraphics[width=12cm ]{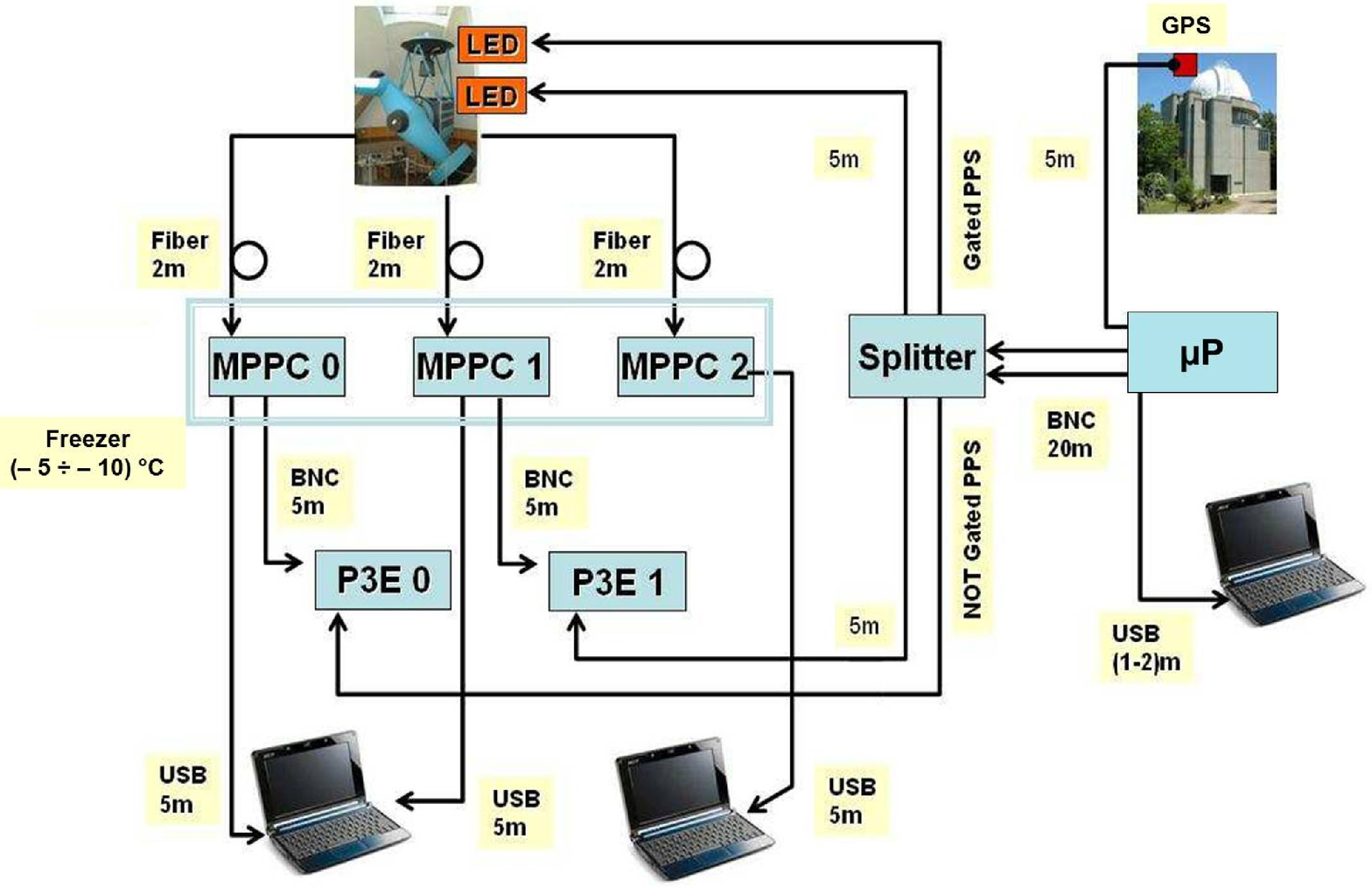}
\caption{General block diagram of the electronic chain mounted at the telescope. The GPS antenna is located outside the dome.}
 \label{figBlEl}
  \end{figure*}

 Our system is composed by three MPPC modules with active area of 1$\times$1 mm$^2$,  pixel size of 50$\times$50 $\mu$m$^2$ and  wavelength range is 320 -- 900 nm with  peak sensitivity at 440 nm. 
The selected pixel size has two consequences. First, a good linear dynamics in counts up to 100 photons simultaneously on the detector surface. Second, the PDE maximum will reach about 50\%.
One detector (MPPC0) is used to observe the target. The second and the third modules 
(MPPC1 and MPPC2) are used to check the transparency of the sky and the telescope
tracking stability during the data acquisition done by MPPC0.  All detectors are controlled by two netbook computers running Windows XP operating system. A third netbook  controls the time acquisition via a GPS device.
Fig. \ref{figBlEl}  shows the general block diagram of our electronic chain. 
 In the present configuration the light from the telescope arrives on  each detector through a plastic optical fiber (600 $\mu$m core diameter).

An appropriate control circuit on each MPPC module monitors the temperature and modifies the bias voltage applied to the sensor itself in order to keep the gain constant against any  temperature variation that could affect the measure especially in the case of long duration measurements.
The electronic noise for each MPPC module is expressed in terms of counts for the whole sensor (i.e. for all the 400 pixels).
To reduce this noise the detectors are kept inside a commercial freezer, where
 within two hours each MPPC module reaches a stable temperature value
  which must be higher than  $-10^{\circ}$C for reliability of the electronics mounted on  board of the MPPC module. 
 At  running conditions  the temperature of our MPPC0 module is $-5.2^{\circ}$C and  the average  count rate is  27.2  counts ms$^{-1}$  with standard deviation of 6.6 counts ms$^{-1}$.
 
The electronics inside each MPPC module
can generate three types of output: analog, discriminated and pulse count via USB interface.
The software provided by Hamamatsu with the MPPC module allows to adjust both the threshold applied to the internal discriminator on the analog signal from the sensor and the gate duration in which the discriminated signal is counted. 



The output with pulse counting is able to give measurements with  a minimum time gate of 1 ms: this output is used by the software suite provided by Hamamatsu to record the acquired data.
 The other two available outputs can be eventually  used  by a home made acquisition system;  we decided to use the discriminated one  to increase  the time sensitivity of our photometer.
The discriminated output is generated from the analog signal processed by a comparator that converts it into a digital one which is in TTL standard logic with minimum duration 20 ns. This output can be extracted  via a standard SMA connector for 50 Ohm impedance coaxial cable.

\begin{figure*}
  \centering
   \includegraphics[width=12cm ]{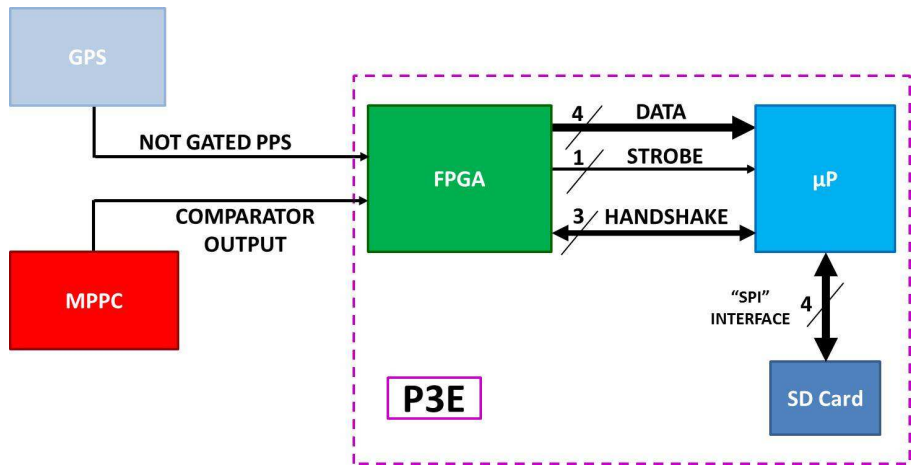}
   \caption{Block diagram of the P3E module.}
         \label{figP3E}
   \end{figure*}

As stated above, the fastest acquisition rate allowed by the software provided by Hamamatsu is 1 ms: we have nearly halved the rate down to 0.55 ms with a dedicated electronic system that we called  P3E (Pulsar Pulse Period Extractor), developed at the Physics Department of La Sapienza University and used during our first measurement of the CRAB Pulsar at the Loiano Observatory where we made most of the tests.
Our P3E electronics process the discriminated signal received from an MPPC module and exploits the possibility to set the threshold level by using the software running on the dedicated netbook computer. The data recording device is an SD card, FAT32 formatted. The block diagram of  an P3E module is shown in  Fig. \ref{figP3E}. 
We use two P3E units, one for the target (P3E0), and one for the sky (P3E1).
Both recording systems (Hamamatsu software and our P3E) are simultaneously active and their outputs are sent to separate files.

During the data acquisition run it is possible to maximize the count rate in the target fine centering phase thanks to the possibility -included in the Hamamatsu software suite- of displaying 
 the sampled data  on the PC monitor in a graphic window. 
 Immediately after the data acquisition, to confirm the correctness of the pointed target,  a quick-look data check can be performed by Fast Fourier Transform (FFT) and autocorrelation analysis using a software like the package MATLAB.
    
\begin{figure*}
  \centering
   \includegraphics[width=12cm ]{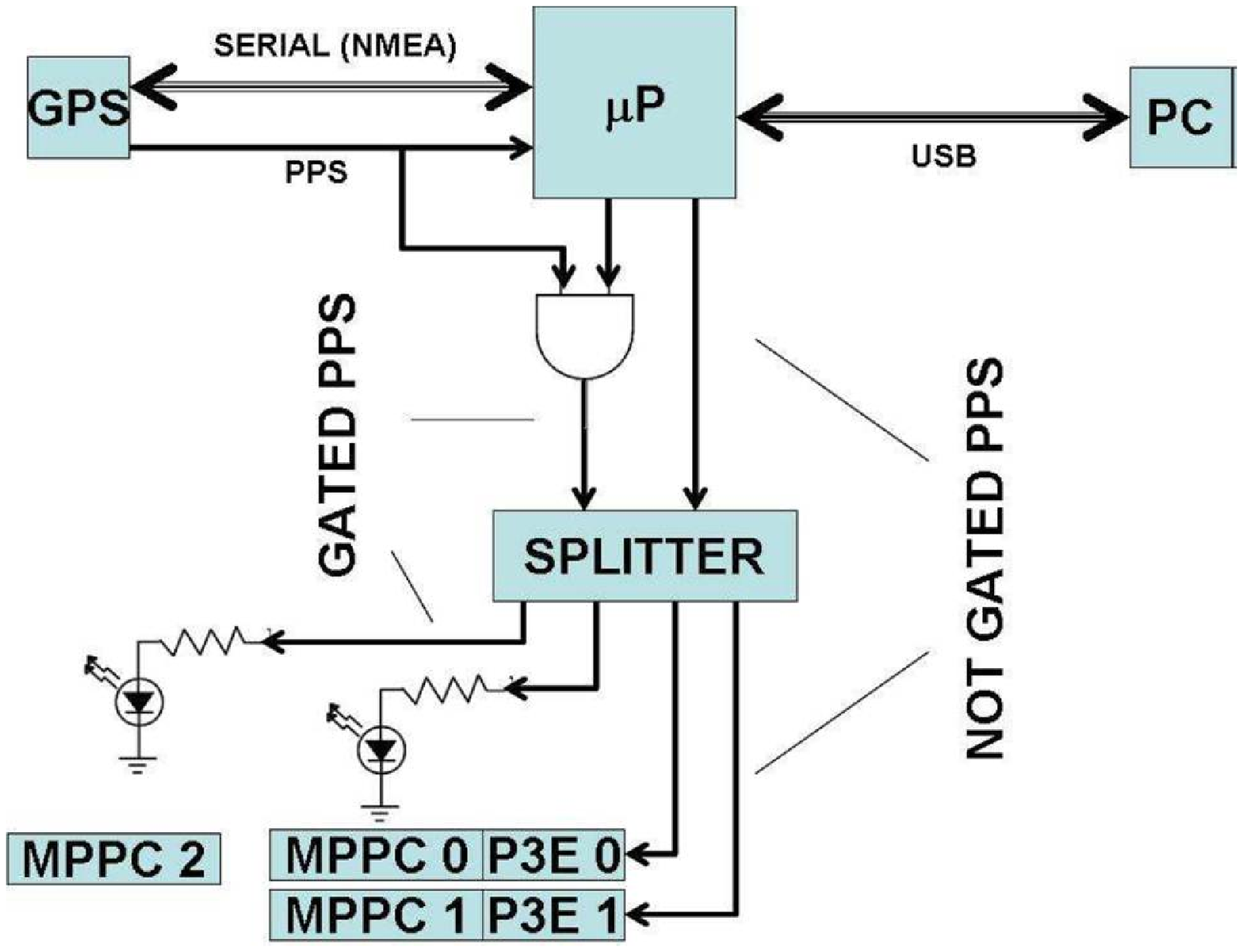}
   \caption{Block diagram of the GPS chain devoted to generate reference time markers on the raw data acquired by the photometer.}
         \label{figGPS}
   \end{figure*}

The Universal Time of the Data Acquisition System is given by a commercial GPS unit (garmin-18x LVC) 
\footnote{http://www8.garmin.com/manuals/GPS18x\_TechnicalSpecifications.pdf }, whose antenna must be located outside the dome and which send the information in National Marine Electronics Association (NMEA\footnote{http://www.nmea.org}) 0183  standard format.
 Fig. \ref{figGPS}  shows the block diagram of the GPS chain devoted to send reference time markers to the acquisition system.

The GPS  sends  through  two  cables, a pulse every second (PPS signal) and the relative information string with serial protocol.
 The cables are connected to a dedicated microprocessor unit (ARDUINO 2009 board, $\mu$P in 
  Fig. \ref{figBlEl}  and \ref{figGPS})  linked to a dedicated netbook: a client software written in C\#  allows the operator to give easily commands to the microprocessor in order to generate a GATED PPS signal, starting from the original PPS signal (hereafter NOT GATED PPS). The same code writes a
  file  containing the date and Universal Time
relative to each PPS  sent in the GATED PPS line. This last  allows one pulse to pass at the starting time and one at the end of each acquisition period. 
The GATED PPS  is sent, through an appropriate line driver  and a  coaxial cable, to a dedicated module close to the telescope,  indicated as Splitter in Fig. \ref{figBlEl}  and \ref{figGPS}.
 
In order to have an optical temporal  marker the output pulses drive two LEDs located near the front end of the optical fibers on the focal plane.
 One LED is used both for MPPC0 and MPPC1, the other one is used for MPPC2.
Being an optical pulse, the time marker is simultaneously present also on the discriminated signals  available from each MPPC module and  therefore it will be also detected by our P3E units. The LEDs are switched on for 0.5 s only at the start and at the end of the acquisition run, and are off otherwise, so they do not produce any background light.
 
  The total delay between the rising edge of the PPS signal at the GPS antenna and  the discriminated output from MPPCs is about 320 ns. The rising edge  of the GPS pulse has a temporal accuracy of $\pm \, 1\, \mu$s.

The NOT GATED PPS signal arrives through a separate cable to the Splitter and the outputs are distributed to each P3E electronic unit to start the internal Finite State Machine, developed using a FPGA (Field Programmable Gate Array), to integrate inside a programmed time window the discriminated signal generated by the relative MPPC module. From each P3E the processed data are sent to a corresponding Microcontroller unit (ARDUINO MEGA 2560  board) that interfaces the mass storage unit. 
  For each detector, the output is a two column ASCII file, containing  a sequential index and the counts in the time bin.

   We have checked both in laboratory and at the telescope, the constancy and accuracy of the time step provided by the Hamamatsu hardware.    



\section {Field tests} \label {field}

 We made the first instrumental tests at the 50 cm telescope
of the University La Sapienza at Vallinfreda \footnote{http://astrowww.phys.uniroma1.it/nesci/vallin.html } and then at the 152 cm Cassini telescope of Bologna  Astronomical Observatory.

 Pointing of the target is performed with the main CCD instrument (BFOSC) of the Cassini telescope permanently mounted on-axis. A flip-mirror located before the focal plane can redirect the light of the central part of the field of view toward our MPPC0 and MPPC1. The sky and the target signals are collected by two optical fibers distant 17 mm apart: at the focal plane each fiber covers 10 arcsec diameter, so the photometry is little sensitive to the seeing conditions.  The typical sky count rate is 5 counts ms$^{-1}$, definitely lower than the electronic dark count rate. 
A third optical fiber, mounted on an independent probe on the focal plane, monitors a reference star. 

\begin{figure}
  \centering
   \includegraphics[ height=8cm ]{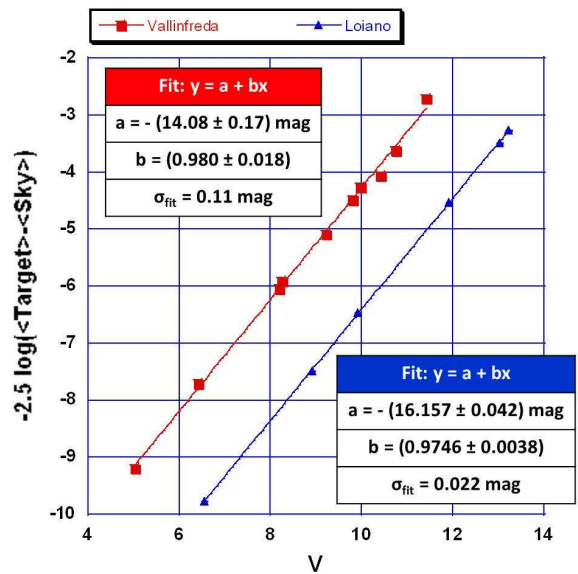}
   \caption{
Observed net count rate in a Pogson law-like scale, as a function of visual magnitude $V$ of the source. 
}
  \label{figCrate}
   \end{figure}

 We checked the overall efficiency and linearity of the instrumental response  between March and July   2010. 
The results  obtained from stars of different magnitudes are plotted in
 Fig. \ref{figCrate}.
 The magnitudes  of the stars were verified by classical photometry obtained during the same nights with BFOSC.
The data were reduced with standard IRAF procedures \footnote{IRAF  is  distributed by  the NOAO which is operated by AURA under contract  with NFS. }. 
  
\begin{figure}
  \centering
   \includegraphics[width=8cm]{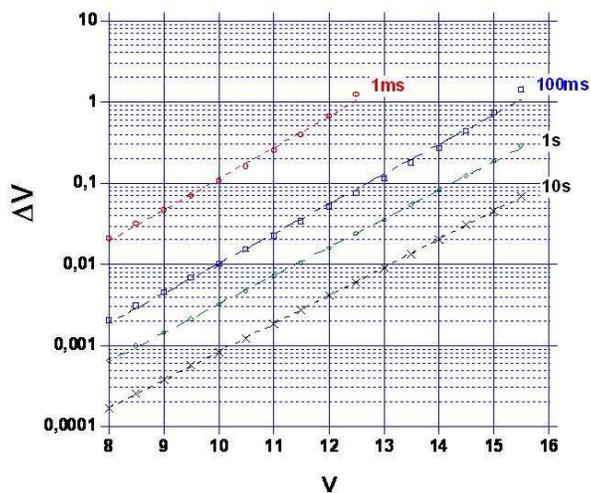}
   \caption{
Expected minimum detectable magnitude variation of a source $\Delta V$ as a function of the source magnitude $V$, for different time gate duration, at the Loiano telescope.
}
\label{figMag}
\end{figure}

 After  having verified the general linearity we used the subsequent runs to study the efficiency of
 the system  in detecting the  flux variability.
According to the noise counts referred before, it is possible to detect faint sources (about 16 mag) with 1 ms integration time and a signal to noise ratio (S/N) $\sim$1 with the Loiano telescope. 
For this telescope we report in Fig. \ref{figMag} the expected sensitivity in magnitude ($\Delta V$) as a function of the visual magnitude $V$ for different  MPPC integration gate length from 1 ms up to 10 s.
For a star of magnitude $ V$ the typical net count  were computed from the fit of Fig. \ref{figCrate}; the $\Delta V$ was obtained  by using  a Pogson law-like formula: 

\begin{equation}
\Delta V =  -2.5\ \log[ 1- \Delta{\rm(NC) / NC} ]
\end{equation} 
 
 where NC is the net count (signal minus background), $\Delta$(NC) is 3 $\sigma$ ~of the typical total background count.
 
 
\section{ Observational test: the Crab pulsar } \label{crab}

On February 5, 2011 we have observed with the 
Cassini telescope the Crab Pulsar
for 3300~s with 0.55~ms (P3E0) and 1~ms (MPPC0) sampling in good photometric 
conditions (seeing $\sim$~1.5~arcsec).

\begin{figure}
\centering
\includegraphics[width=8cm]{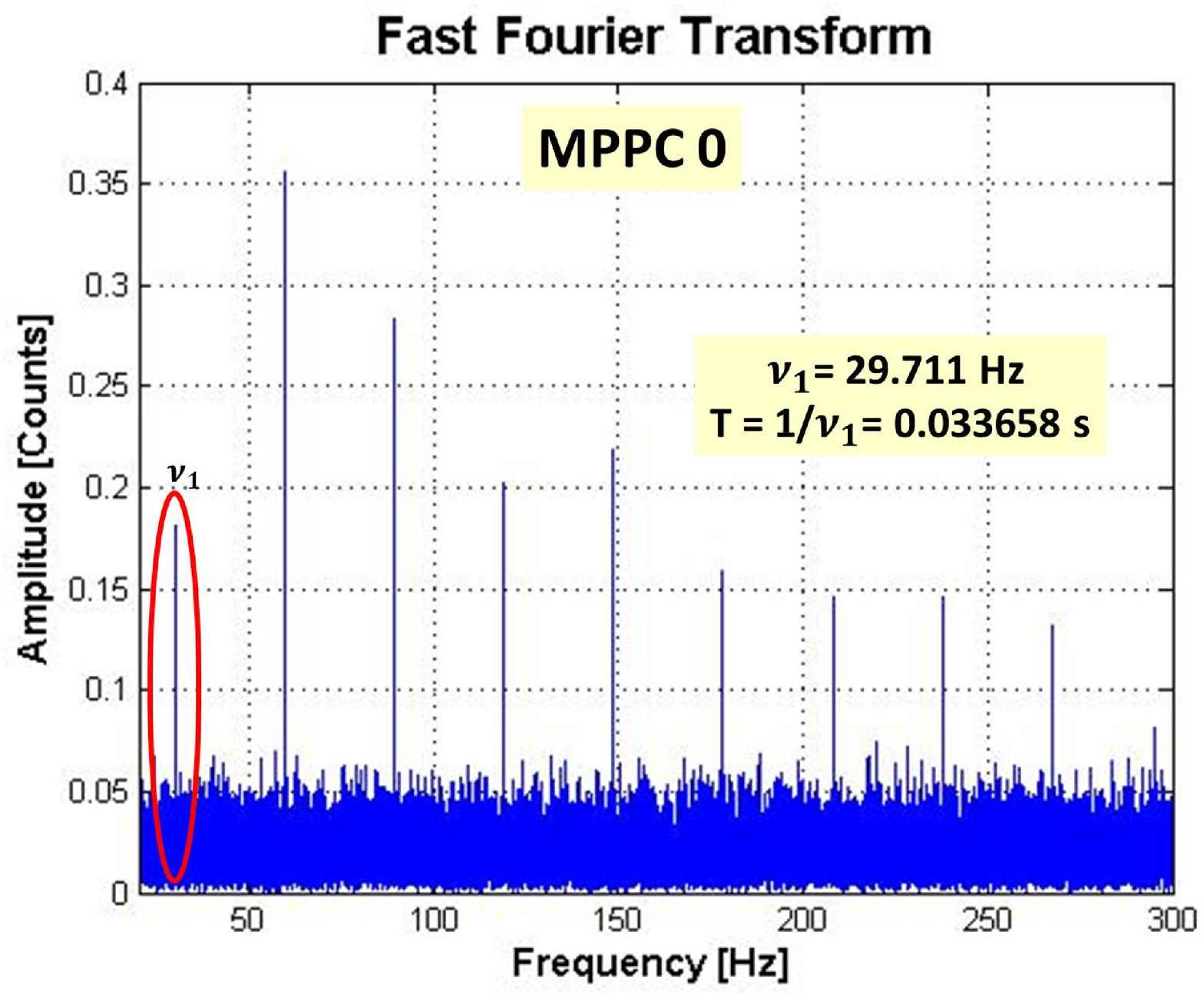}   
\caption{ FFT applied to the first 100 s of raw data acquired by MPPC0. }
\label{figfft_mppc}
\end{figure}

\begin{figure}
\centering
\includegraphics[width=8cm]{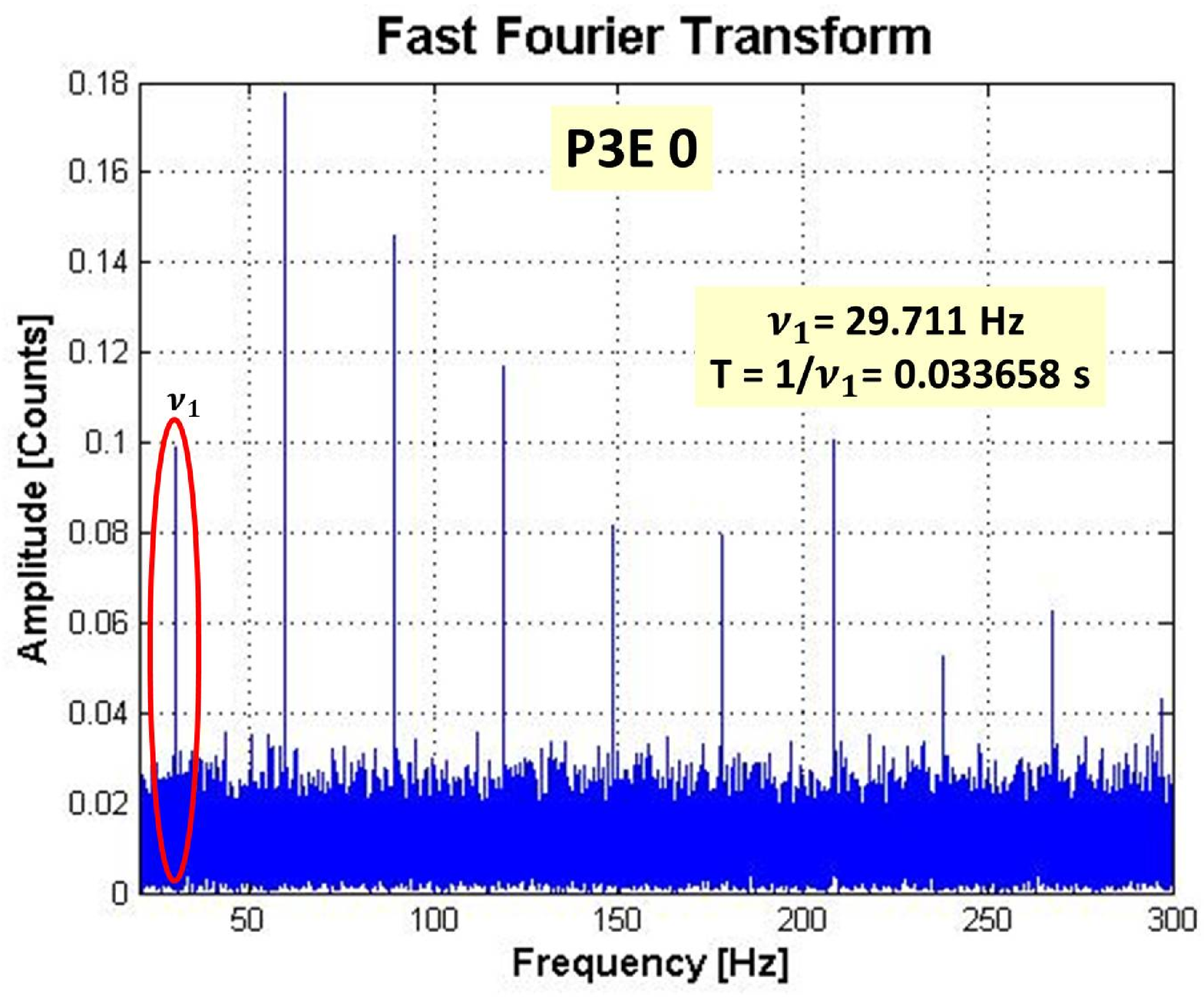}   
\caption{ FFT applied to the first 100 s of raw data acquired by P3E0. }
\label{figfft_p3e0}
\end{figure}

 The autocorrelation function applied to a section of the raw data  emphasized 
the  presence of a  periodic signal close to  the expected value.
A period search with FFT applied to a short section of the total acquisition 
time (only 100~s), provided a period of 0.033658~s  for  both  MPPC0 and P3E0, 
in fair agreement with the expectation from the Jodrell Bank Observatory 
ephemeris (0.033652394~s) \footnote{http://www.jb.man.ac.uk/pulsar/crab.html}
(see Fig. \ref{figfft_mppc} and \ref{figfft_p3e0}).
As expected the same analysis applied to the whole raw data set improves the signal to 
noise ratio, the number of the detectable harmonics and the agreement of the 
period (0.033654734~s for MPPC0 and 0.033655303~s for P3E0) with the expected  one. 

To apply the heliocentric correction we used the Xronos software from HEASARC
\footnote {http://heasarc.nasa.gov/docs/xanadu/xronos/xronos.html} using
the task $earth2sun$.
 This package does not take into account the topographic correction for the observatory coordinates and therefore it is not suited for a check of the phase shift between radio and optical signal.  This check was not an aim of the  present work, which is  mainly devoted to the technical details of the acquisition system.
 At the same time the various phase differences reported in literature are of the order of hundreds
  $\mu$s, below our temporal resolution, e.g.  (Barbieri et al. 2009, Oosterbroek et al. 2008).   

\begin{figure*}
  \centering
 \includegraphics[width=\textwidth  ]{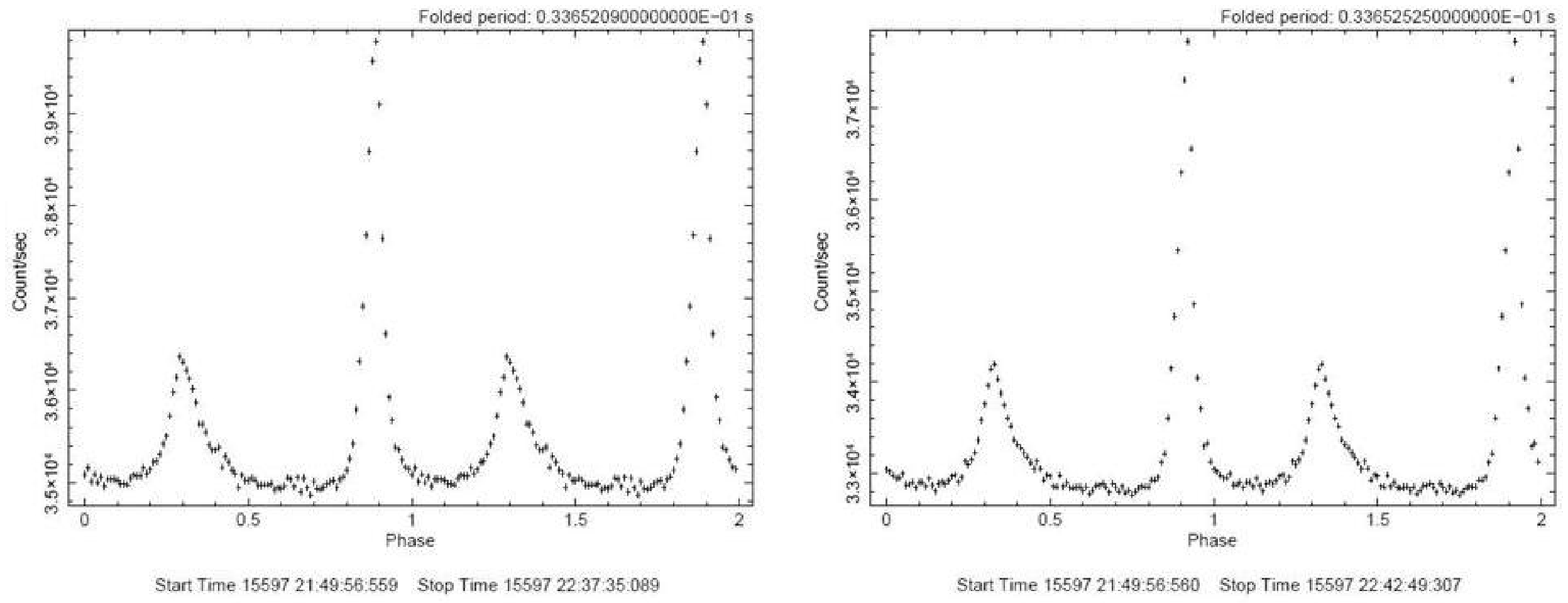}  
\caption{Crab Pulsar light curves folded by the Xronos task $efold$ for 
  MPPC0 (left) and P3E0 (right). The data were prevously barycentered with 
  the task $earth2sun$. The phases are arbitrary and not related to the 
  Jodrell Bank Observatory ephemeris.}
         \label{figlicur}
   \end{figure*}

We then  used the task $efsearch$ to determine the best fitting period.
We obtained T$_{Crab}$ = 0.033652090~s for MPPC0 and T$_{Crab}$ = 0.033652525~s 
for P3E0, respectively. 
Finally we built the Crab light curve with the task $efold$ and the results are reported in Fig. \ref{figlicur}
which shows that shape and flux ratio between the primary and secondary pulse are in good agreement with literature data, see e.g. Zampieri et al. (2011) and references therein.
The phase shift between  MPPC0 and P3E0 is 0.04,  corresponding to about 1.3 ms, comparable with the accuracy of  the MPPC0 module sampling rate.

As a check, we applied the same procedures of periodicity search ($earth2sun, efsearch, efold$) also to the raw data coming from the other two MPPC modules (sky and control star): no  sign of periodicity was detected,  confirming therefore the validity of 
our result on the Crab Pulsar.


\section{Conclusions and developments } \label{conclusion}

Our analysis demonstrated that our instrumentation can detect the Crab 
Pulsar signal with a 1.5~m class telescope.

With the present configuration a quick look of the data  performed on a section ($\sim$~100~s) already  allowed to detect the presence of periodic signal with a  period close to the expected one.

Immediate future developments  will  mainly concern the improvement of the acquisition system.
~~{\it i)} ~The first aim is to increase the S/N ratio and the simplicity of the system.
Hamamatsu has recently commercialized a new MPPC module with the 
same physical  characteristics  but with a built-in Peltier cooler  at $-10^{\circ}$C fixed temperature and dark level  lower than  10 count ms$^{-1}$.
Such module will avoid the need for a freezer and the need for an optical
fiber between the telescope focal plane and the sensor. 
~~{\it ii)}  ~The second point aims to obtain a more accurate determination of the starting
and ending time of the acquisition run: we are testing a new software to drive 
the GATED PPS signal with a burst of {\it n} PPS instead of an individual one.
~~{\it iii)} ~The third point concerns the sampling step: the speed limit is not given by 
the SiPM module but by the data recording device SD card. The next electronics development 
will try to improve the limit by modifying both the software and the hardware components.

\acknowledgments
~The authors would like to thank Bologna Observatory for the logistic support 
and the technical assistance during the observations. 
~The University La Sapienza supported the project.
~This research has made use of SIMBAD database operated at CDS, Strasbourg, 
France.
~F. M. would like to express his personal thanks to Engineer M. Aversa of the 
Italian branch of Hamamatsu Photonics for the technical help and the fruitful 
discussions about the MPPC module characteristics applied in this unusual 
astronomical field.  
~The authors thank the anonymous referee  for the careful work and  detailed criticism  that 
helped us to improve the paper.



{\it Facilities:}  \facility{Bologna Cassini Telescope (BFOSC)}.

\clearpage
\end{document}